\documentclass[superscriptaddress]{article}
\usepackage{amssymb,amsthm,amsmath,amsfonts,physics,algorithm2e,float}
\usepackage[a4paper, total={7in, 9in}]{geometry}
\usepackage[normalem]{ulem}
\usepackage{ulem,soul,enumerate}
\usepackage{graphicx}
\usepackage[pdftex,dvipsnames,usenames]{xcolor}
\usepackage[colorlinks=true,urlcolor=blue,citecolor=blue,linkcolor=blue]{hyperref}
\usepackage{authblk}
\usepackage[sorting=none,backend=biber,style=nature]{biblatex}
\usepackage{multicol}
\usepackage{caption}
\usepackage{subcaption}
\usepackage[dvipsnames]{xcolor}
\usepackage{comment}
\usepackage{siunitx}
\DeclareUnicodeCharacter{0301}{*************************************}
\newcommand{\minus}{\scalebox{0.75}[1.0]{$-$}}

\begin{document}
\title{Wavevector-resolved polarization entanglement from radiative cascades}
\author[1]{Alessandro Laneve\thanks{equal contribution}\thanks{alessandro.laneve@uniroma1.it}}
\author[1]{Michele~B.~Rota$^*$}
\author[1]{Francesco~Basso Basset$^*$}
\author[1]{Mattia~Beccaceci}
\author[1]{Valerio~Villari}
\author[2]{Thomas~Oberleitner}
\author[3]{Yorick~Reum}
\author[2]{Tobias~M.~Krieger}
\author[3]{Quirin~Buchinger}
\author[2,4]{Saimon~F.~Covre~da~Silva}
\author[3]{Andreas Pfenning}
\author[5]{Sandra~Stroj}
\author[3]{Sven~Höfling}
\author[2]{Armando~Rastelli}
\author[3]{Tobias~Huber-Loyola}
\author[1]{Rinaldo~Trotta\thanks{rinaldo.trotta@uniroma1.it}}
\affil[1]{Dipartimento di Fisica, Sapienza Università di Roma, Piazzale Aldo Moro 5, I-00185 Roma, Italy}
\affil[2]{Institute of Semiconductor and Solid State Physics, Johannes Kepler University, Altenbergerstraße 69, Linz 4040, Austria}
\affil[3]{Technische Physik, University of Würzburg, Am Hubland, D-97074 Würzburg, Germany}
\affil[4]{Universidade Estadual de Campinas, Instituto de F\'{i}sica Gleb Wataghin, 13083-859 Campinas, Brazil}
\affil[5]{Forschungszentrum
Mikrotechnik, FH Vorarlberg, Hochschulstr. 1, A-6850 Dornbirn, Austria}

\date{\today}
\maketitle

\begin{abstract}
The generation of entangled photons from radiative cascades has enabled milestone experiments in quantum information science with several applications in photonic quantum technologies. Significant efforts are being devoted to pushing the performances of near-deterministic entangled-photon sources based on single quantum emitters often embedded in photonic cavities, so to boost the flux of photon pairs. The general postulate is that the emitter generates photons in a nearly maximally entangled state of polarization, ready for application purposes. Here, we demonstrate that this assumption is unjustified. We show that in radiative cascades there exists an interplay between photon polarization and emission wavevector, strongly affecting quantum correlations when emitters are embedded in micro-cavities. We discuss how the polarization entanglement of photon pairs from a biexciton-exciton cascade in quantum dots strongly depends on their propagation wavevector, and it can even vanish for large emission angles. Our experimental results, backed by theoretical modelling, yield a brand-new understanding of cascaded emission for various quantum emitters. In addition, our model provides quantitative guidelines for designing optical microcavities that retain both a high degree of entanglement and collection efficiency, moving the community one step further towards an ideal source of entangled photons for quantum technologies.
\end{abstract}
\begin{figure}[H]
 \centering
\includegraphics[width=0.95\linewidth]{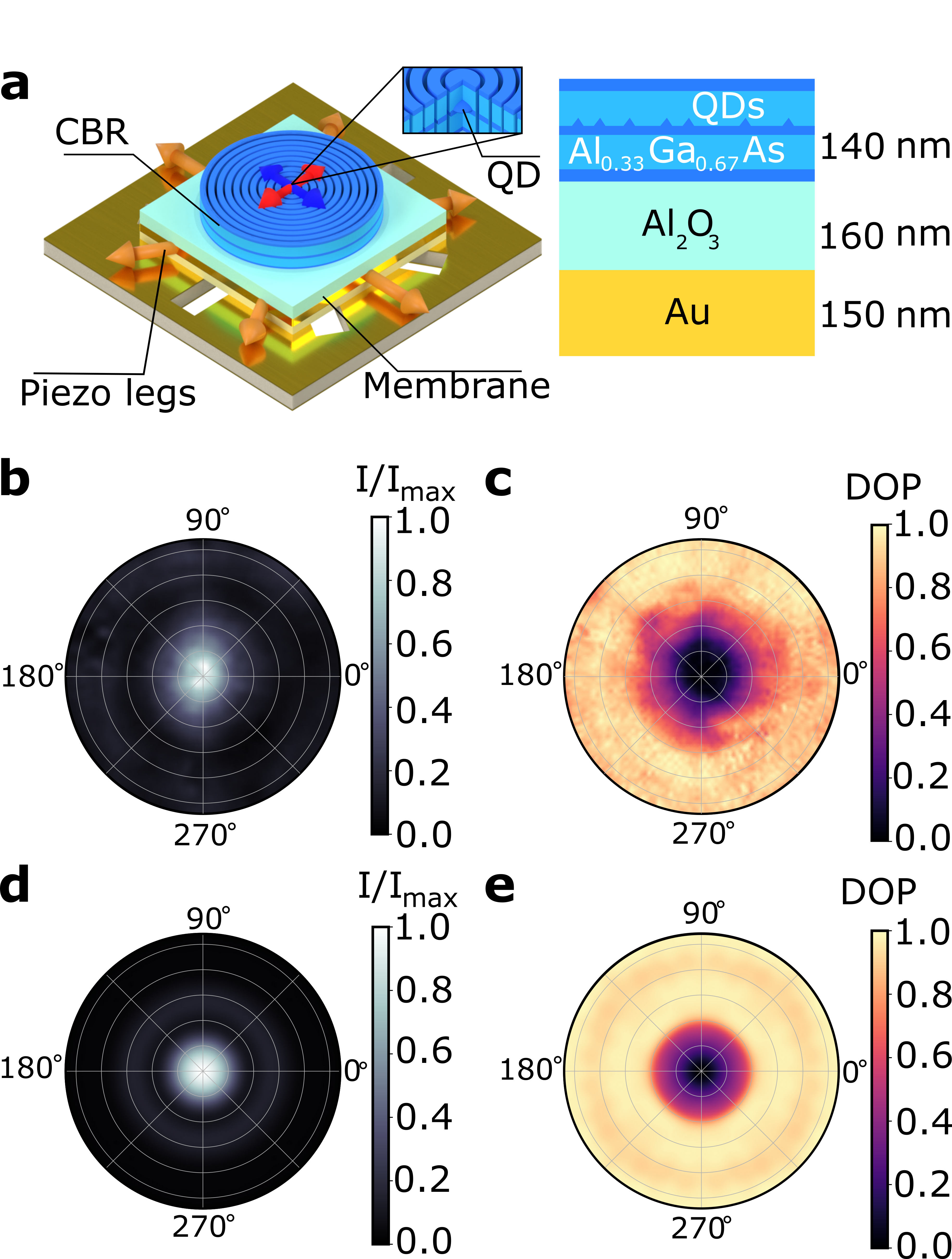}
\caption{\textbf{Far-field emission and degree of polarization for quantum dots in bullseye cavities.} \textit{\textbf{a} Sketch of the CBR sample. A detailed description of the sample features is provided in the SI.
    \textbf{b} The experimental far-field intensity distribution for a GaAs QD in a CBR cavity and \textbf{c} corresponding DOP distribution along the radial axis. Circles represent $10^\circ$ increments in the zenithal angle.
    The DOP is computed from the Stokes parameter sampled on the far-field radiation of the QD, which was observed through BFP imaging. \textbf{d} Intensity for the CBR as computed from FDTD simulations. \textbf{e} DOP for the CBR as computed from FDTD simulations. The full Stokes parameters distributions are reported in SI. 
    }}
    \label{fig:experimental_dop}
\end{figure}

\begin{multicols}{2}
The employment of radiative cascades from single emitters (atoms) has been at the core of the first demonstrations of Bell inequality violation \cite{freedman1972experimental,aspect1981experimental}. 
Since the outset of experimental quantum information more “practical” solid-state emitters are extensively being exploited as quantum light sources. 
Radiative cascades are found in a wide variety of modern, atom-like emitters such as epitaxial \cite{benson2000regulated,akopian2006entangled} and colloidal \cite{yin2017bright,utzat2019coherent} quantum dots (QDs), NV-centers in diamonds \cite{togan2010quantum, ruf2021quantum}, molecules \cite{norden2018entangled, rezai2019polarization}, and defects in 2D materials \cite{he2016cascaded, tonndorf2017single}.
In a cascaded emission, polarization entanglement of photons can directly stem from basic symmetries and selection rules \cite{yang1950selection}, and a two-step cascade should provide maximally polarization-entangled photon pairs; however, it is important to note that this is the case only if the pairs are collected at suitable wavevectors. This notion dates even back to the first demonstrations of photonic non-locality \cite{clauser1969proposed, shimony1971foundations}, where it was predicted - but not experimentally demonstrated - that the quantum correlation between photons generated from an atomic radiative cascade depends on the light collection angle. Thus, polarization entanglement of photon pairs is intertwined with their emission pattern, which can be generally assimilated to that of two oscillating dipoles \cite{scully1999quantum}. Nonetheless, in the successive studies and experiments that employed radiative cascades, from the oldest \cite{freedman1972experimental,aspect1981experimental} to more recent ones \cite{benson2000regulated,akopian2006entangled}, this feature has been generally overlooked or neglected. The reason is straightforward: in both \cite{clauser1969proposed} and \cite{shimony1971foundations}, analytical calculations on atomic systems show that the practical consequences of this effect only become relevant for wide angles of collection. 
We recognize that this also holds for solid-state emitters for which, despite the presence of a preferential emission direction corresponding to the main quantization axis, the outcome of the radiative cascade for small collection angles is analogous. 
Emission angles away from the surface normal are instead hard to access due to total internal reflection \cite{benisty1998method}. 
That said, neglecting the angular dependence of the photonic state generation becomes untenable when we consider the current need to develop quantum light sources with unprecedented brightness. 
  On-demand emitters, such as QDs, have been recently embedded in different families of cavities \cite{gerard1998enhanced,hennessy2007quantum,lodahl2015interfacing,dousse2010ultrabright,liu2019solid,wang2019demand}, all with the same purpose: increasing the extraction efficiency of photons while preserving quantum correlations. It should be noted that many
claims of record results (in terms of brightness joined to
entanglement) are based on the assumptions that the
sampled set of photons is actually representative of all the
photons emitted. Indeed, most of these structures funnel light into a small observation cone to collect as much signal as possible. Thus, it is reasonable to expect that a polarization-wavevector correlation appears at yet accessible collection angles. Nevertheless, to the best of our knowledge, no experiments have ever reported on such an effect, despite the incredible efforts currently underway to optimise the performances of deterministic entangled photon sources. 
Here, we demonstrate experimentally the existence of a strong interplay between the degree of entanglement and light emission angle. We also developed a model for the two-photon state generated by a radiative degenerate two-level cascade that includes the angular dependence of entanglement, with which we start our discussion.\\
\indent In typical radiative cascade processes here analyzed, entangled photon emission arises from the presence of a doubly-excited state, which can be written in terms of the single particle total angular momentum projection along the optical axis of collection, $\ket{J_{\boldsymbol{z},1}}\ket{J_{\boldsymbol{z},2}}$. In systems like QDs, where the $\boldsymbol{z}$-axis is the main confinement direction, the doubly excited state, neglecting its dark components and fine structures, is given by \cite{juska2015conditions}:
\begin{equation}
   \ket{\Psi}=\frac{1}{\sqrt{2}} \bigg(\ket{1}\ket{-1}+\ket{-1}\ket{1}\bigg).
   \label{eq:excitations_state}
\end{equation}
\begin{figure*}[h!t!]
\includegraphics[width=\linewidth]{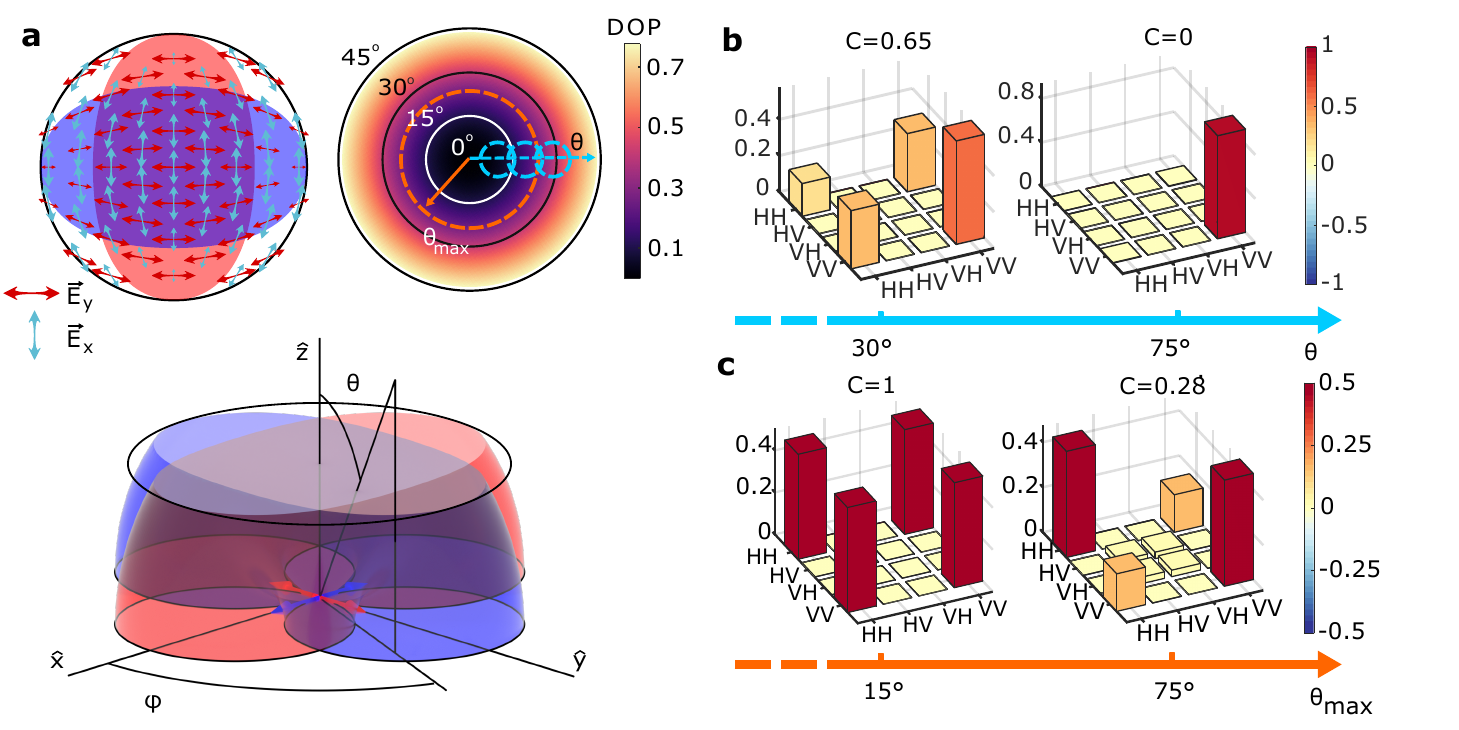}
\caption{\textbf{Theoretical predictions for the polarization state of photons emitted by a radiative cascade.} \textit{\textbf{a} Bottom panel: sketch of the two emitting dipoles (blue and red arrows) with the respective equipotential surfaces of the radiation intensity.
Top left panel: planar cut of the total emission pattern along with the direction of the electrical field oscillation for both dipole radiations, highlighting the intensity mismatch as well as the dependence of the orientation of both fields on the emission angle. Top right panel: overall degree of polarization (DOP), defined as the square root of the squared sum of the Stokes parameters (see the SI), as a function of the angles $\phi$ and $\theta$.
The dashed orange circle indicates the considered far-field region centred to the quantum emitter, corresponding to an emission angle up to $\theta_{max}$. The dashed blue circles indicate the different $\theta$s that select specific wavevectors.
\textbf{b} Real part of the two-photon entangled state as a function of $\theta$. The value of the concurrence for both angles is also reported. \textbf{c} Real part of the two-photon entangled state as a function of $\theta_{max}$. The value of the concurrence for both angles is also reported.}}
\label{fig:dipole_emission}
\end{figure*}
This doubly-excited state radiatively decays following the selection rule $\Delta J=\pm1$, and generating a two-photon state that, given the conservation of angular momentum to the circularly polarized photons $\{\ket{R},\ket{L}\}$, is maximally entangled in the polarization degree of freedom:
\begin{equation}
\begin{split}
     \ket{\psi}=\frac{1}{\sqrt{2}}(\ket{R}\ket{L}+\ket{L}\ket{R})=\\
     =\frac{1}{\sqrt{2}}(\ket{H}\ket{H}+\ket{V}\ket{V})=\ket{\phi^+},
\end{split}
    \label{eq:bellstate}
\end{equation}
 where $\ket{\phi^+}$ is a maximally entangled Bell state in the linear polarization basis $\{\ket{H},\ket{V}\}$.
The two-photon state resulting from a degenerate radiative cascade is generally regarded as the one in Eq.~\eqref{eq:bellstate}, both for actual atoms \cite{freedman1972experimental,aspect1981experimental} and QDs \cite{benson2000regulated,akopian2006entangled,hudson2007coherence}. 
However, this is true only when we consider photons propagating with $\boldsymbol{k}=\boldsymbol{z}$, where $\boldsymbol{k}$ is the single-photon wavevector and $\boldsymbol{z}$ is the confinement direction.
The behaviour for $\boldsymbol{k}\neq\boldsymbol{z}$ is dictated by the selection rules and how they link the polarization state of a single photon generated from one transition (with a specific variation of total angular momentum) to its emission direction.
The general $\{\boldsymbol{k}\}$-dependent state of polarization  can be written in terms of polar coordinates:
\begin{equation}
   \ket{P^{\pm}_{\boldsymbol{k}}}=\gamma_{\theta}^{\pm}(\boldsymbol{k})\ket{\theta}+\gamma_{\phi}^{\pm}(\boldsymbol{k})\ket{\phi},
   \label{eq:polarization_wavevector}
\end{equation}
where $\{\ket{\theta},\ket{\phi}\}$ represent polarization vectors along the zenithal angle $\theta$ and the azimuthal angle $\phi$, and the $\gamma^\pm_{\theta,\phi}(\boldsymbol{k})$ are the components along $\theta$ and $\phi$ of the $\gamma^\pm(\boldsymbol{k})$ vector, which depends on the dipole moment of the $\ket{\pm1}$ to ground state transition (see the Methods section for its expression).
We can also understand why this $\boldsymbol{k}$-dependency is needed in terms of angular momentum conservation: the projection of the total angular momentum of the excited state is transferred to the total angular momentum of the photon, which corresponds to its polarization only when it is emitted at $\boldsymbol{k}=\boldsymbol{z}$. For arbitrary $\boldsymbol{k}$s, the polarization only corresponds to the spin angular momentum state of the photon \cite{shimony1971foundations}. 
  In order to explicitly compute the $\boldsymbol{k}$-dependence of $\gamma^\pm(\boldsymbol{k})$, we start by considering an emitter in vacuum.\\ 
  \indent The two transitions leading to Eq.~\eqref{eq:excitations_state} can be pictured
as two radiating dipoles, which we assume as independent, and whose fields add incoherently.
 The intensity pattern we calculated by our model is shown in Fig.~\ref{fig:dipole_emission}\textbf{a} for both dipole emitters, together with the total degree of polarization (DOP) as a function of the light emission angle. It is clear that as we step away from the condition $\boldsymbol{k}=\boldsymbol{z}$, the emission features a net degree of polarization, due to an unbalanced mixing of the two dipole fields: at large angles, light will come mostly from one of the two dipoles, thus the emission will be endowed with the corresponding polarization.
 It is important to understand that the wavevector-polarization correlation that we have discussed for single photon states 
 has also profound consequences on the degree of entanglement of photon pairs generated during the radiative decay of the state described in Eq.~\eqref{eq:excitations_state}.
By applying the $\{\boldsymbol{k}\}$-dependent selection rules to the total angular momentum state of Eq.~\eqref{eq:excitations_state}, the polarization state changes from the case $\boldsymbol{k}=\boldsymbol{z}$ reported in Eq.~\eqref{eq:bellstate} to the more general expression:
\begin{equation}
\begin{split}
     \ket{\psi}=A \int_{\boldsymbol{k},\boldsymbol{k}'}d \boldsymbol{k}d \boldsymbol{k'} C_{\boldsymbol{k}} C_{\boldsymbol{k}'} \ket{\boldsymbol{k},\boldsymbol{k}'}\big(\ket{P^+_{\boldsymbol{k}}}\ket{P^-_{\boldsymbol{k}'}}+\\+\ket{P^-_{\boldsymbol{k}}}\ket{P^+_{\boldsymbol{k}'}}\big),
    \end{split}
\label{eq:two-photon_ketstate_general}
\end{equation}
where $A$ is a normalization factor, $\boldsymbol{k}$ and $\boldsymbol{k}'$ represent the independent modes in which the two photons can be emitted, and $C_{\boldsymbol{k}/\boldsymbol{k}'}$ represents the probability density for a photon to be emitted in mode $\boldsymbol{k}/\boldsymbol{k}'$.
With Eq.~\eqref{eq:two-photon_ketstate_general} at hand, it is instructive to visualize how the entangled two-photon state looks like for an emitter in vacuum, for which we can use a plane-wave decomposition to find an analytic form of the $\gamma$ coefficients as a function of $\theta$ and $\phi$.
For a wavevectors' pair, identified by the angles $\{\theta,\phi,\theta',\phi'\}$, the two-photon polarization state can be written as:
\begin{equation}
\begin{split}
        \ket{\psi(\theta,\theta')}=\frac{\cos(\theta)\cos(\theta')\ket{\theta \theta'}+\ket{\phi \phi'}}{\sqrt{1+\cos(\theta)^{2}\cos(\theta')^{2}}},
\end{split}
\label{eq:two-photon_ketstate_dipole}
\end{equation}
since cylindrical symmetry cancels any dependence of the polarization coefficients from the angles $\{\phi, \phi'\}$.
If a fixed observation point is set, so that $\theta=\theta'$, the two-photon state results more and more polarized by increasing $\theta$, leading to a complete disruption of entanglement, as highlighted in Fig.~\ref{fig:dipole_emission}\textbf{b}. 
We can easily understand why this is the case because at very large angles only one of the two cascades is collected. This introduces which-path
information that reduces the degree of entanglement down to zero.
A more realistic light-gathering scenario is instead obtained by integrating over a solid angle of collection $\Omega$, which, because of cylindrical symmetry, we can describe in terms of $\theta_{max}=\Omega/2\pi$, i.e., the maximum $\theta$ for which the signal is collected.
Figure~\ref{fig:dipole_emission}\textbf{c} shows how the density matrix of the photon pair changes as we widen the collection aperture, leading again to an overall entanglement deterioration (see Supplementary Information\textemdash SI\textemdash for the analytical expression as a function of $\theta_{max}$ and its complete derivation).
So far, we have assumed that the bright part of the doubly-excited state is described by Eq.~\eqref{eq:excitations_state}, which works well for QDs when photon pairs are generated during the biexciton (XX) - exciton (X) radiative cascade \cite{benson2000regulated}.
It also works for other solid-state quantum emitters subject to similar confinement conditions \cite{utzat2019coherent,he2016cascaded}. 
However, it is important to recognize that for real atoms with spherical symmetry, the emission intensity is isotropic and the loss of entanglement arises from another physical phenomenon. As the observation aperture is increased, radiation from further allowed transitions is collected, so that the effective atomic state is \cite{shimony1971foundations} 
\begin{equation}
   \frac{1}{\sqrt{2}}\big(\ket{1,-1}+\ket{0,0}+\ket{-1,1}\big)
   \label{eq:ketstate_atom}
\end{equation}
and the degree of entanglement is deteriorated because light from all three possible transitions is collected as the acceptance aperture is increased.
Nonetheless, our theory can be extended also to this particular case by simply using the formalism we have developed and Eq.~\eqref{eq:ketstate_atom} as the initial state.\\
\indent To bring the above theoretical discussion into a real-life framework, we move from an emitter in vacuum to one in a solid-state device. 
We chose a state-of-the-art entangled photon source: 
 a GaAs QD in an Al$_{0.33}$Ga$_{0.67}$As \cite{huo2013ultra} membrane with a bottom broadband oxide-metal mirror embedded in a circular Bragg resonator (CBR) \cite{liu2019solid}.
 This specific structure has the advantage of a broadband collection enhancement, resulting in an extraction efficiency of $\eta \approx 0.7$ for both the transitions of interest and can be integrated on top of a micromachined piezoelectric actuator to induce anisotropic strain \cite{rota2024source}, as depicted in Fig.~\ref{fig:experimental_dop}\textbf{a}. Strain engineering is used to erase the Fine Structure Splitting between the transitions associated with the two dipoles of the emitter and generally push the degree of entanglement close to unity when light from $\boldsymbol{k}=\boldsymbol{z}$ is collected \cite{benson2000regulated,basso2021quantum,trotta2016wavelength,rota2024source,huber2017highly}.
Differently from the vacuum case, the far-field emission of the emitter now depends on the micro-cavity features and we cannot use a simple plane-wave expansion to find an expression for the $\gamma^\pm(\boldsymbol{k})$ coefficients.
Rather, we have to resort to numerical simulations that critically depend on the symmetry, the geometrical parameters of the cavity as well as the position of the quantum emitter, as discussed in previous works \cite{rickert2019effective}.
For this reason, we assessed the polarization-resolved emission profile of our source with dedicated experiments and simulations. To gain information on the degree of polarization from light as a function of light emission angle (as shown in Fig.~\ref{fig:dipole_emission}\textbf{a} for the emitter in vacuum), we assembled a Fourier microscopy setup (details in the Methods section and SI) to record the back 
\noindent focal plane (BFP) image of the QD emission, which contains $\boldsymbol{k}$-space information of the emitted light \cite{lieb2004single}. In this way, we can perform high-resolution polarimetry of the radiation far-field \cite{lieb2004single,osorio2015k,li2015efficient,fons2018all}.
 In Fig.~\ref{fig:experimental_dop}\textbf{b} and \ref{fig:experimental_dop}\textbf{c}, we report the experimental intensity distribution for the single photon emission of the X to ground state transition and the corresponding DOP distribution. While more than $90\%$ of the light is contained within $40^\circ$, the DOP noticeably starts deviating from zero at around $20^\circ$ of angular aperture (the acceptance angle of an off-the-shelf multimode optical fibre) and reaches a DOP$=0.5$ at around $\theta\approx30^\circ$. The pattern is analogous for the XX transition, although it presents some differences that we discuss in the SI.
 The main features of the experimental results are found as well in the corresponding finite-difference time-domain (FDTD) simulations of the sample emission, performed by the Lumerical software, both for the intensity distribution, the DOP (Fig.~\ref{fig:experimental_dop}\textbf{d} and Fig.~\ref{fig:experimental_dop}\textbf{e}), and the entire set of Stokes parameters (see the SI). Even if there is an excellent qualitative agreement between the experiments and the simulations, the latter predict a narrower cone of emission and a steeper change of DOP. 
 These differences can be ascribed to the fact that the simulations were performed considering the nominal CBR parameters, and there is a strong dependence on the exact geometry of the micro-resonator (which cannot be easily assessed experimentally without causing damages that would alter its optical properties), as discussed in the SI.
  That said, the strong similarity between Fig.~\ref{fig:dipole_emission}\textbf{a} and Fig.~\ref{fig:experimental_dop}\textbf{c} suggests that there must exist a strong interplay between the degree of entanglement and light emission angle. To experimentally demonstrate this effect,
 we collect light at a specific $\boldsymbol{k}$ region, by setting up a spatial filter on the BFP image of the QD emission (see Methods).
 We assess the degree of entanglement of the $\boldsymbol{k}$-selected photon pairs by separating XX and X photons with two volume Bragg gratings and performing quantum state tomography.
 Multimode optical fibres are then employed to guide light to the single photon detectors used to record the coincidences needed to reconstruct the two-photon density matrices.
\begin{figure*}[t!]
    \centering
\includegraphics[width=\linewidth]{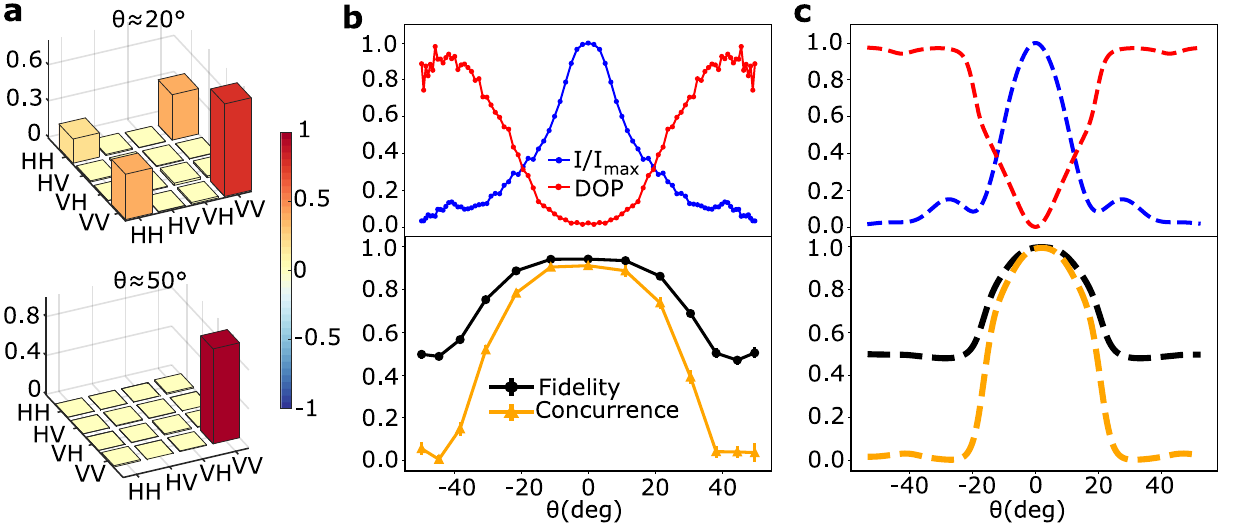}
\caption{
\textbf{Two-photon state and entanglement figures of merit as a function of the main collection angle} \textit{ \textbf{a} Two examples of experimental density matrices for different main collection angles $\theta$. Due to collection from a polarized wavevector region, the two-photon state has a well-defined polarization and entanglement is lost. \textbf{b} Intensity and DOP profiles as a function of $\theta$, extrapolated by the BFP measurements. They are reported with the corresponding fidelity of entanglement to the target state $\ket{\phi^+}$ and concurrence, computed as a function of the main collection angle $\theta$ and for a small $\boldsymbol{k}$ integration range (selected by a $20$ $\mu$m pinhole on the BFP, see the SI). For the experimental data, we average over the azimuthal angle $\phi$ to obtain an average profile for radiation intensity and DOP as a function of only $\theta$. The same quantities  are reported in \textbf{c}, intensity and DOP resulting from far-field FDTD simulations of the emission together with fidelity and concurrence computed on two-photon states obtained by inserting the simulated fields in Eq.~\eqref{eq:two-photon_ketstate_general}.
 Error bars are computed assuming Poissonian distributions of coincidences and they are remarkably small due to the high number of recorded events.}}
\label{fig:ent_vs_theta}
\end{figure*}
We initially compare the density matrices sampled by scanning along the diameter of the emission cone, with a fixed narrow angular aperture, similarly to the blue circles in Fig.~\ref{fig:dipole_emission}\textbf{a}. Figure~\ref{fig:ent_vs_theta}\textbf{a} reports two examples of two-photon density matrices collected respectively at $\theta=20^\circ$ and $\theta=50^\circ$. 
Alongside, in Fig.~\ref{fig:ent_vs_theta}\textbf{b}, we report two main entanglement figures of merit, i.e., concurrence and fidelity, as a function of $\theta$. The degree of entanglement dramatically decreases when photons are collected from a highly polarized wavevector region, as highlighted by the corresponding DOP profile. 
Since both photons in the cascade have a defined polarization, the which-path information of the decay channel is revealed and entanglement is lost. This is also extremely clear inspecting the density matrix recorded for $\theta=50^\circ$, showing the complete absence of the coherence terms and a fully polarized two-photon state.
 \begin{figure}[H]
    \centering
\includegraphics[width=\columnwidth]{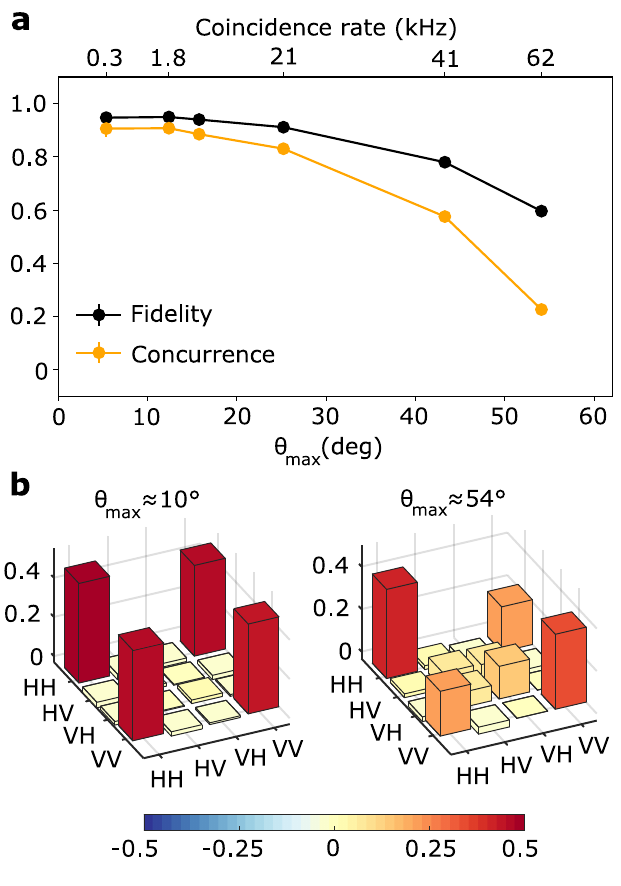}
\caption{
\textbf{Two-photon state and entanglement figures of merit as a function of the collection aperture.} \textit{ \textbf{a} Fidelity to the $\ket{\phi^+}$ state and concurrence as a function of different sizes of the collection region centred around the main propagation axis. Data are shown as a function of the corresponding collection half-angle $\theta_{max}$. \textbf{b} Experimental density matrices corresponding to different values $\theta_{max}$, showing how the two-photon polarization state follows the description of Eq.~\eqref{eq:densmat_theo}. 
Error bars are computed assuming Poissonian distributions of coincidences (they are remarkably small due to the high number of recorded events).}}
\label{fig:ent_vs_thetamax}
\end{figure}
\noindent  These experimental results look in excellent qualitative agreement with the simple theory discussed for the two dipoles in vaccum (see Fig.~\ref{fig:dipole_emission}\textbf{b)}, the main difference being the angle at which the degree of entanglement starts to drop.
By exploiting FDTD simulation, we are also able to predict the evolution of the entangled state for a QD in a CBR cavity. Specifically, we can compute the polarization vector of light for any wavevector and both the possible transitions, i.e., radiating dipoles.
 Then, we insert the normalized polarization vector in the corresponding term in Eq.~\eqref{eq:two-photon_ketstate_general}, choosing a suitable polarization basis. Given this state, we compute quantum correlations as a function of the wavevector or range of wavevectors (for details see the SI). We report the results of this analysis in Fig.~\ref{fig:ent_vs_theta}\textbf{c}, showing an excellent qualitative agreement with experimental data. 
 Quantitatively, the main difference is related to the FDTD simulations expecting a narrower cone of emission and a steeper change of the DOP and the degree of entanglement, a discrepancy once again related to the use of the nominal CBR parameters in the simulations (see the discussion above and the SI). We performed a similar analysis also on lower extraction efficiency structures, GaAs QDs in planar membranes with and without DBR reflectors. As anticipated and further reported in the SI, in these cases the effect can be present but much less pronounced due to the limited access to wider and more polarized emission angles.\\ \indent We performed additional experiments to investigate how the two-photon state changes as a function of the numerical aperture in the collection optics. We do that by using spatial filters with different dimensions in the BFP, similarly to the orange circle of Fig.~\ref{fig:dipole_emission}\textbf{a}. Fig.~\ref{fig:ent_vs_thetamax}\textbf{a} highlights a rapid decrease of quantum correlation between the emitted photons as the collection half-angle $\theta_{max}$ increases.
It is important to emphasize that for this type of experiment - in which we collect light with spatial filters always centred at the maximum of the light intensity profile in the BFP -  the drop in the degree of entanglement is not connected to the net DOP (i.e., DOP integrated over the whole emission), which always remains below about $8\%$. 
This experimental evidence is particularly relevant to exclude that the phenomenon that we observe here could be related to possible deviations from the cylindrical symmetry of the device and/or bad positioning of the QD in the centre of the cavity. These effects would indeed result in an overall sizeable DOP \cite{peniakov2024polarized}, that we instead do not observe theoretically nor experimentally (see Fig.~\ref{fig:experimental_dop}).
By inspecting the density matrices of Fig.~\ref{fig:ent_vs_thetamax}\textbf{b} we can also exclude the entanglement drop we observe in Fig.~\ref{fig:ent_vs_thetamax}\textbf{a} could be attributed to the imperfect overlap of orthogonally polarized modes \cite{larque2009optimizing}. This should result in simple decoherence \cite{larque2009optimizing} and would not justify the additional terms of state mixing that can be observed in the matrix with the larger collection aperture.
More specifically, starting from Eq.~\eqref{eq:two-photon_ketstate_general}, we can obtain the following description of the two-photon polarization density matrix:
\begin{equation}
\begin{split}
       &\rho(\theta_{max})=\\
       &=\big(1\minus p_1(\theta_{max})\minus p_2(\theta_{max})\minus p_3(\theta_{max})\big)\ket{\phi^+}\bra{\phi^+}+\\
       &+\frac{p_1(\theta_{max})}{2}\big(\ket{HH}\bra{HH}+\ket{VV}\bra{VV}\big)+\\&+p_2(\theta_{max})\ket{\psi^+}\bra{\psi^+}+\\
       &+\frac{p_3(\theta_{max})}{2}\big(\ket{HV}\bra{HV}+\ket{VH}\bra{VH}\big),
\end{split}
\label{eq:densmat_theo}
\end{equation}
which contains the incoherent mixing of different terms, including $\ket{\psi^+}=\frac{1}{\sqrt{2}}(\ket{HV}+\ket{VH})$, weighted by the functions $p_1(\theta_{max})$, $p_2(\theta_{max})$, and $p_3(\theta_{max})$ that depend on $\theta_{max}$ and that can be numerically computed from the $\boldsymbol{k}$-dependent polarization vectors for each dipole far-field emission. The presence of a mixing with the $\ket{\psi^+}$ state in the experimental density matrix (see Fig.~\ref{fig:ent_vs_thetamax}\textbf{b} for $\theta_{max}=54^\circ$), which once again follows our theory (see Fig.~\ref{fig:dipole_emission}\textbf{c}),  clearly rules out the imperfect overlap between two orthogonally polarized modes as the only cause of this effect. 
  Eq.~\eqref{eq:densmat_theo} can be analytically calculated for the dipole-in-vacuum case, and in the SI we discuss how it can be effectively generalized to cavity-embedded emitters.
Finally we point out that Eq.~\eqref{eq:densmat_theo} offers a valuable tool in combination with FDTD simulations to include the expected degree of entanglement in the design and optimization of a photon pair source.\\
\indent Deterministic quantum emitters promise to be the next generation of quantum light sources and their properties have been extensively studied, benchmarked and engineered. 
In this work, we add a novel and previously overlooked aspect to the picture: in radiative cascades, different electronic transitions can generate radiation with distinct angular distributions.
 This feature produces a wavevector-polarization correlation affecting the entanglement of emitted photon pairs, a general effect that had never been witnessed. We managed to observe it for the first time in a cavity-embedded QD, thanks to the light-funneling cavity design enhancing the emergence of the phenomenon. 
We demonstrate here that the wavevector-polarization correlation is a fundamental phenomenon that arises from the basic properties of the emitter and its nanophotonic environment; this effect may pose a trade-off between the degree of polarization-entanglement and brightness of the source in state-of-the-art photonic cavities designed to boost the flux of entangled photons. Fig.~\ref{fig:ent_vs_thetamax}\textbf{a} shows that the maximum degree of entanglement is obtained only by filtering considerably over the wavevector range in the BFP, a technique which comes at the cost of the rate of collected photon pairs. 
To provide numbers that better explain the existence of this trade-off, we have removed the set-up used to image the back focal plane and directly coupled the QD emission in a single mode fibre (using a set-up that minimizes losses and it is usually used in quantum communication experiments \cite{basso_basset_quantum_2021,rota2020entanglement}) as well as in a multi-mode fibre. 
Whereas collecting $100\%$ of the QD signal in multi-mode fibres delivers a $0.60(2)$ fidelity, the entanglement fidelity using single-mode fibres is $0.94(1)$ while retaining about $30\%$ of the coincidence rate. This behaviour also concurs to explain why this phenomenon has been neglected so far, as state-of-the-art quantum optics experiments are performed using single-mode fibers.
That said, we strongly believe that it is possible to devise strategies that allow to preserve brightness together with the the maximum degree of entanglement. For instance, one could employ
the model we report here to design photonic cavities that cancel out the wavevector-polarization correlation while keeping high photon-extraction efficiency. 
On the other hand, it may be possible to fully recover entanglement post-emission, by polarization-sensitive manipulation of the signal wavefront.
In conclusion, our results unveil novel fundamental features of cascaded quantum emitters, which can impact other crucial characteristics in the development of nanophotonic systems, such as photon indistinguishability. In addition, our work points the way to engineer light-matter interaction to achieve the ultra-bright source of entangled photons required for real-life applications of quantum communication.
\section*{Methods}
\subsection*{Experimental setup}
The far-field images were acquired through a Fourier microscopy setup: light emitted by the QD is collected and collimated by an objective having NA$=0.81$, which fixes the numerical aperture of the whole system and allows for the collection of photons emitted at a maximum angle of $\sim54$° to the direction orthogonal to the surface. Afterwards, a system of two lenses, $L_1$ and $L_2$, suitably positioned along the photons path and featuring focal length $f_1=100$ cm and $f_2=40$ cm, respectively, allows to project the back focal plane image of the QD emission on a CCD camera sensor of $1340\times100$ square pixels each of dimensions $20\;{\mu}m\times20\;{\mu}m$. The different contributions due to XX-X and X-ground transitions could be distinguished by filtering the corresponding emission lines using volume Bragg gratings \cite{glebov2012volume} and suitable lowpass and longpass filters.
To select specific $\boldsymbol{k}$s regions, an additional telescope made by two $f=3$ cm lenses is inserted after $L_1$, within its focal length. It generates another back focal plane image in the path towards the CCD, where by inserting a spatial filter wavevector can be selected. Thanks to removable pinholes of different sizes interposed between the two lenses in their focal points, and the ability to move them along the plane orthogonal to the signal propagation direction, we can select specific wavevectors of the radiation and different angular ranges of wavevectors.
To measure the two-photon density matrix, XX and X are separately extracted using suitable volume Bragg gratings and they undergo a quantum state tomography \cite{james2001measurement}.
Further detailed information about the setup and the source are reported in the SI.

\subsection*{Theoretical model for the dipoles in vacuum}
Optical transitions are ruled by the minimal-coupling Hamiltonian \cite{scully1999quantum}.
Here, we treat single photon emission from a QD as a consequence of an exciton decay from an excited state to the ground state in a two-level system. Since we coherently pump the XX state, and the splitting between the heavy (HH) and light (LH) bands is in the order of tens of meV, a reasonable assumption is to consider only the creation of bright HH excitons. Minor changes are sufficient to also include HH-LH mixing and are left for future studies. Therefore, we will call excited and ground states respectively the conduction ($\ket{s}\ket{\pm}$) and valence ($\mp\frac{1}{\sqrt{2}}(\ket{p_{x}} \pm i\ket{p_{y}})\ket{\pm}$) band states in the $s$-shell of the QD, where $\ket{s}$ represents the $s$-type orbital, while $\ket{p_x}$ and $\ket{p_y}$ represent the $p$-orbitals, composed with the spin state $\ket{\pm}$.
An accurate description of the output photonic state must account for the transition coupling to different possible field modes. The system can be described by the Hamiltonian $\mathbf{\mathcal{H}}=\mathbf{\mathcal{H}}_{el} + \mathbf{\mathcal{H}}_{at} + \mathbf{\mathcal{H}}_{int}$, where $\mathbf{\mathcal{H}}_{el}=\sum_{\boldsymbol{k}}\hbar\omega_{\boldsymbol{k}}(\hat{a}_{\boldsymbol{k}}^{\dag}\hat{a}_{\boldsymbol{k}}+\frac{1}{2})$ and $\mathbf{\mathcal{H}}_{at}=\frac{1}{2}\hbar\omega\sigma_{z}$ \cite{scully1999quantum}. The interaction term of the system Hamiltonian $\mathbf{\mathcal{H}}_{int}=-e\boldsymbol{r}\cdot\mathbf{E}$, that regulates radiative transitions, can be written in terms of creation and annihilation operators both of photons and atomic-like excitations. Then, decomposing the electrical field in a mode basis $\{\boldsymbol{k}\}$,  $\mathbf{E}=\sum_k A_{\boldsymbol{k}} \hat{\mathcal{E}}_{\boldsymbol{k}}(\hat{a}_{\boldsymbol{k}}+\hat{a}^{\dag}_{\boldsymbol{k}})$ where $\hat{\mathcal{E}}_{\boldsymbol{k}}$ is the polarization vector for the mode $\boldsymbol{k}$ and the dipole transition term $e\boldsymbol{r}=\sum_{ij} \xi_{ij}\ket{i}\bra{j}=\sum_{ij} \boldsymbol{\xi}_{ij}\sigma_{ij}$, we obtain:
\begin{equation}
    \mathbf{\mathcal{H}}_{int}=\sum_{ij}\sum_{k}\hbar A_{\boldsymbol{k}}\boldsymbol{\xi}_{ij}\cdot\mathcal{\hat{E}}_{\boldsymbol{k}} \sigma_{ij}(\hat{a}_{\boldsymbol{k}}+\hat{a}^{\dag}_{\boldsymbol{k}})
\end{equation}
 The information about how the photon polarization is distributed in space is encoded inside the $\gamma^{\pm}({\boldsymbol{k}})=A_{\boldsymbol{k}} \xi_{ij}$ term that we employ in Eq.~\eqref{eq:polarization_wavevector}. 
 The "wavefunction" of a photon generated by a dipole emitter can be written in terms of plane waves \cite{scully1999quantum}:
 \begin{equation}
   \centering
   \Psi^{\pm}(\mathbf{r},t)=\sum_{k} \sqrt{\frac{\hbar\omega_{\boldsymbol{k}}}{2\epsilon_{0}V}}\frac{\gamma^{\pm}(\boldsymbol{k})}{r}\frac{e^{i(\boldsymbol{k}\cdot\boldsymbol{r}-\omega_{\boldsymbol{k}}t)}}{\Delta_{\boldsymbol{k}}+i\frac{\Gamma_{i}}{2}}
\end{equation}
being $\hbar\omega_{\boldsymbol{k}}$ the emitted photon energy (which differs from the gap between ground and excited state by a quantity $\Delta_{k}$) $\Gamma_{i}$ the decay rate of the i-th excited state, and $V$ an arbitrary quantization volume.
In case the emitter is in vacuum, the $A_{\boldsymbol{k}}$ dependence is trivial and, being $(0,\hat{\theta},\hat{\phi})$ a general expression for the field polarization versor $\hat{\mathcal{E}}_{\boldsymbol{k}}$, we derive a functional form of the emitted photon polarization by carrying out the scalar product between the dipole matrix element and $\hat{\mathcal{E}}_{\boldsymbol{k}}$ in the oscillator strength definition. 
\begin{equation}
    \ket{P^{\pm}(\boldsymbol{k})}=\mp\frac{\Pi e^{\pm i\phi}}{\sqrt{2}}(\cos{\theta}\ket{\hat{\theta}} \pm i\ket{\hat{\phi}})
    \label{eq:photon_pol}
\end{equation}
which is what one should expect from a $\Delta J =\pm 1 $ transition, where we denote $\Pi=\bra{s}x\ket{p_{x}}=\bra{s}y\ket{p_{y}}$. The relative phase between $\ket{p_{x}}$ and $\ket{p_{y}}$ is transmitted to the photon polarization, which turns out to be circularly polarized only if we harvest light travelling along $\hat{z}$ direction ($\theta=0^{\circ}$), while it becomes linearly polarized for wider zenithal collection angles.
Assuming that the photon emission direction does not correlate with the transition process properties, we can build the two-photon state as a composition of two coupled $\Delta J =\pm 1 $ cascade transitions, i.e.,
\begin{equation}
\begin{split}
    \ket{\Psi_{1,2}}=\\
    =A\sum_{\boldsymbol{k}_1,\boldsymbol{k}_2}C(\boldsymbol{k}_{1})C(\boldsymbol{k}_{2})\big(\ket{\boldsymbol{k}_{1},P^{+}(\boldsymbol{k}_{1})}\ket{\boldsymbol{k}_{2},P^{-}(\boldsymbol{k}_{2})} +\\
    +\ket{\boldsymbol{k}_{1},P^{-}(\boldsymbol{k}_{1})}\ket{\boldsymbol{k}_{2},P^{+}(\boldsymbol{k}_{2})}\big)
\end{split}
    \label{two-photon_state_sum}
\end{equation}
where $A$ is a normalization factor, while $C(\boldsymbol{k}_{1})$ and $C(\boldsymbol{k}_{2})$ represent the probability for photon 1 or 2 to be emitted along a specific direction. We can derive Eq.~\eqref{eq:two-photon_ketstate_general} by projecting the spatial component of the joint photonic wavefunction onto two specific $\boldsymbol{k'}_{1}$ and $\boldsymbol{k'}_{2}$, i.e., $\bra{\boldsymbol{k'}_{1}}\bra{\boldsymbol{k'}_{1}}\ket{\Psi_{1,2}}$=$\ket{\Psi_{1,2}}\delta(\boldsymbol{k}_{1}-\boldsymbol{k'}_{1})\delta(\boldsymbol{k}_{2}-\boldsymbol{k'}_{2})$. We can then re-write Eq.~\eqref{two-photon_state_sum} as:
\begin{equation}
\centering
\begin{split}
     \ket{\Psi_{1,2}(\boldsymbol{k}_{1},\boldsymbol{k}_{2})}=\\
     =A\cdot C(\boldsymbol{k'}_{1})C(\boldsymbol{k'}_{2})
\big(\ket{P^{+}(\boldsymbol{k'}_{1})}\ket{P^{-}(\boldsymbol{k'}_{2})}+\\
+\ket{P^{-}(\boldsymbol{k'}_{1})}\ket{P^{+}(\boldsymbol{k'}_{2})}\big)
\end{split}
\label{cesemoquasi}
\end{equation}
By substituting Eq.~\eqref{eq:photon_pol} in Eq.~\eqref{cesemoquasi}, and defining the constant $A$ from the state normalization, we can obtain the functional form of the entangled two-photon state in Eq.~\eqref{eq:two-photon_ketstate_dipole}.

\printbibliography

\section*{Data availability}
The data that support the results of this paper are available
from the corresponding authors upon reasonable request. 

\section*{Code availability}
The code supporting the analysis reported in this paper is available
from the corresponding authors upon reasonable request.

\section*{Acknowledgments}
This project has received funding from the European Union’s Horizon 2020 research and innovation program under Grant Agreement no. 899814 (Qurope) and No. 871130 (Ascent+), and from the QuantERA II program that has received funding from the European Union’s Horizon 2020 research and innovation program under Grant Agreement No 101017733 via the project QD-E-QKD and the FFG grant no. 891366. The authors also acknowledge support from MUR (Ministero dell’Università e della
Ricerca) through the PNRR MUR project PE0000023-NQSTI, the European Union’s Horizon Europe research and innovation program under EPIQUE Project GA No. 101135288, the Austrian Science Fund (FWF) via the Research Group FG5, I 4320, I 4380, and the Linz Institute of Technology (LIT), the LIT Secure and Correct Systems Lab, supported by the State of Upper Austria.  T.H.L. acknowledges financial support from the German Ministry of Education and Research (BMBF) within the Project Qecs (Förderkennzeichen 13N16272). We thank Paolo Mataloni and Emanuele Pelucchi for useful discussion.

\section*{Authors contribution}
A.L., M.B.R., and F.B.B. contributed equally to the results of the paper.
A.L., M.B., and F.B.B. developed the theoretical model for the polarization state of photon pairs, as well as the corresponding numerical simulations.
A.L., M.B.R., M.B., V.V., and F.B.B. performed the experiment with help from R.T.. M.B.R., M.B., V.V. and F.B.B. designed the experimental setup.
F.B.B and R.T. designed the source setup. 
T.O. and Y.R. provided numerical simulations data to support experimental results with the supervision of A. P..
S.F.C.d.S
and A.R. designed and grew the QD sample. M.B.R.,
T.M.K, Q.B., and S.S., designed and processed the cavity
of the QD with the supervision of R. T, A.R, T. H. L. and S. H.. A.L. wrote the manuscript, with feedback from all the authors. 
The project was conceived and coordinated by R.T..

\section*{Competing interests}
The authors declare no competing
financial or non-financial interests.
\end{multicols}

\end{document}